\documentclass[twocolumn]{article}

\usepackage{PRIMEarxiv}

\usepackage{booktabs} 
\usepackage[utf8]{inputenc} 
\usepackage[T1]{fontenc}    
\usepackage[hidelinks]{hyperref}
\usepackage{url}            
\usepackage{booktabs}       
\usepackage{amsmath}
\DeclareMathOperator*{\argmax}{arg\,max}

\usepackage{braket}
\usepackage{amsfonts}       
\usepackage{nicefrac}       
\usepackage{microtype}      
\usepackage{lipsum}		
\usepackage{graphicx}
\usepackage{doi}
\usepackage{multirow}
\usepackage{subcaption}
\usepackage{wrapfig}
\usepackage{bbm}
\usepackage{tikz}
\usepackage{quantikz}
\usepackage{adjustbox}
\usepackage[square,numbers,comma,sort&compress]{natbib}
\usepackage{placeins}

\usepackage{float}

\usepackage{tikz}
\usetikzlibrary{arrows}
\usepackage{multicol}

\newtheorem{definition}{Definition}

\title{Quantum Neural Networks under Depolarization Noise: \\ Exploring White-Box Attacks and Defenses}

\author{
  David Winderl, Nicola Franco, Jeanette Miriam Lorenz \\
  Fraunhofer Institute for Cognitive Systems IKS \\
  Munich, Germany \\
  \texttt{\{name.middlename.surname\}@iks.fraunhofer.de} \\
}

\begin{document}

\twocolumn[ 
  \begin{@twocolumnfalse} 
  
\maketitle

\begin{abstract}
Leveraging the unique properties of quantum mechanics, Quantum Machine Learning (QML) promises computational breakthroughs and enriched perspectives where traditional systems reach their boundaries.
However, similarly to classical machine learning, QML is not immune to adversarial attacks.
Quantum adversarial machine learning has become instrumental in highlighting the weak points of QML models when faced with adversarial crafted feature vectors. 
Diving deep into this domain, our exploration shines light on the interplay between depolarization noise and adversarial robustness.
While previous results enhanced robustness from adversarial threats through depolarization noise, our findings paint a different picture.
Interestingly, adding depolarization noise discontinued the effect of providing further robustness for a multi-class classification scenario.
Consolidating our findings, we conducted experiments with a multi-class classifier adversarially trained on gate-based quantum simulators, further elucidating this unexpected behavior.
\end{abstract}


\vspace{0.35cm}

  \end{@twocolumnfalse} 
] 

\section{Introduction}
The field of quantum computing has shown fast advancements in recent years, providing the possibility to speed up specific tasks. Hereby a variety of algorithms exists such as Shor's algorithm to factor primes~\cite{shor1994algorithms}, Grover's search to find an element in an unsorted list in $\mathcal{O}(\sqrt{n})$~\cite{grover1996fast} or the HHL-Algorithm to efficiently solve systems of linear equations~\cite{harrow2009quantum}. In this context, quantum machine learning (QML) focuses onto providing the properties and speedups of quantum computing in the field of machine learning~\cite{ventura2000quantum, trugenberger2002quantum, schuld2018supervised, biamonte2017quantum}.
First coined by \citet{KAK1995143}, the term \textit{quantum neural network} (QNN) emerged as an ambitious effort to merge observations from the domain of neuroscience with the intriguing features inherent to quantum computations. Essentially, QNNs represent a specific type of Quantum Machine Learning (QML) models. Inspired by the layered approach of classical neural networks. 

\begin{figure}[hbt!]
    \centering
    \includegraphics[width=\linewidth]{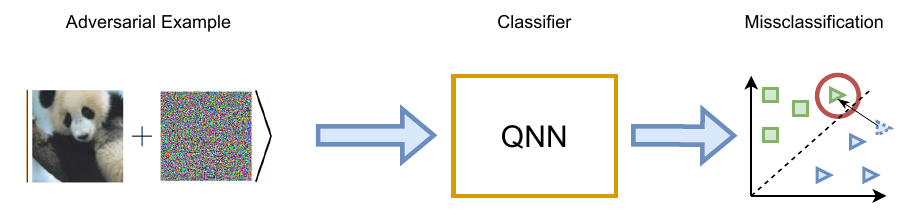}

                

    \caption{Outline of the process used in this study. Overall, the input is perturbed classically and then encoded on the quantum device using amplitude encoding.}
    \label{fig:structure}
\end{figure}
As their classical counterpart, \citet{lu2020quantum} highlighted that QNNs are not immune to adversarial attacks. Adversarial attacks involve carefully designed disturbances that can mislead machine learning models, causing them to produce incorrect predictions. Such attacks have long been a significant issue in traditional machine learning~\cite{biggio2013evasion, szegedy2013intriguing, goodfellow2015explaining, aleks2017deep}. The importance of investigating adversarial attacks on QNNs is reinforced by the \textit{no free lunch theorem}~\cite{poland2020no}. This principle suggests that there is not a single model or algorithm universally optimal for all tasks, hinting at inherent vulnerabilities also in quantum-based models. Hence, finding these weak spots is crucial for the resilience and protection of quantum machine learning systems.
A further obstacle is that current quantum hardware used to execute these models suffers from noise and decoherence times, which is the reason why \citet{preskill2018quantum} coined the term NISQ-Era for the current state in quantum computing. This means, that due the noise, limited connectivity and decoherence times, inherent in quantum devices, current algorithms can only be executed on smaller toy problems, providing a proof of concept, but suffering from not yet being applicable to practical problems.

\paragraph{Our Contribution}
In this work, we explore the effect of noise, specifically depolarization noise, on the robustness of Quantum Neural Networks (QNNs), aiming to minimize compromises on accuracy and training data. We assumed the structure in \autoref{fig:structure} as the most realistic scenario, so the input is perturbed on the classical side and then encoded on the quantum circuit. Furthermore, our research builds on the theoretical insights of \citet{Du_2021}, which suggest that QNNs inherently possess an upper bound of robustness relative to the device's depolarization noise. This upper bound arises because depolarizing noise, viewed as a stochastic Pauli-channel, effectively scrambles the state of the quantum device, thus blurring information. This blurring leads to a natural tradeoff between accuracy and robustness. We evaluate the effectiveness of this phenomenon by adding multiple depolarizing noise channels to a quantum neural network and comparing its performance with other adversarial training techniques commonly used in classical settings.

Our results indicate that depolarization noise does not improve adversarial robustness in multi-class classifiers in realistic settings. Additionally, we observed that increasing the number of classes diminishes both accuracy and overall robustness, with depolarization noise offering no significant enhancement.


\section{Related Work}
Since quantum neural networks are prone to adversarial attacks~\cite{liu2020vulnerability,lu2020quantum},
over recent years, multiple authors have tried to use quantum computing techniques to enhance adversarial robustness for both classical and quantum neural networks. 
As an example, \citet{sahdev2023adversarial} have utilized a quantum phase estimation to provide a quadratic speedup for randomized smoothing. However, because quantum phase estimation is not feasible on NISQ-Hardware and \citet{sahdev2023adversarial} requires a nested quantum phase estimation specifically for quantum neural networks, we did not use this method as a comparison in our studies.
\citet{gong2022enhancing} proposed a vanishing gradient for introducing adversarial noise on a quantum circuit. Nevertheless, this method assumes that the data is already in a quantum state, subject to adversarial perturbation during the execution of the quantum circuit. Since we focused on encoding classical, adversarial perturbed data on the quantum circuit rather than assuming already encoded quantum states, we refrained from including \citet{gong2022enhancing} in our study.
In the context of encoding classical data which is adversarially perturbed, \citet{Weber_2021} have provided a fundamental link between quantum hypothesis testing and binary classification. Their work also shows the connection between classical randomized smoothing and depolarization noise on a quantum circuit. Furthermore during the course of this study, \citet{huang2023certified} have provided a theoretical study on the fact that random noise in the form of rotational gates can improve the overall robustness. Additionally, \citet{berberich2023training} have introduced a method for training weights to increase the Lipschitz bound of quantum neural networks, thereby enhancing their robustness.
Considering specifically noise in quantum neural networks, \citet{wang2023quantumnat} suggests that error mitigation of noise as well as noise aware training helps to improve the overall accuracy of the classifier.  

\section{Background}
\subsection{Quantum information}
The fundamental unit of information in quantum computers is the quantum bit (qubit), which is a superposition of the two quantum-mechanical states: 
\begin{eqnarray}
        \psi =& \alpha \ket{0} + \beta \ket{1}  \quad \text{with}&~{|\alpha|}^2 + {|\beta|}^2 = 1.
    \label{eq:qubit}
\end{eqnarray}
Unlike classical computers that use bits with values of either 0 or 1, quantum computers can simultaneously represent both states. 
This allows for inherent parallel processing in computations. 
However, when concluding a calculation on a quantum computer, the resulting quantum-mechanical state needs to be measured to convert the calculation result into classical bits. 
Due to the superposition, the result of the measurement is not definite, but instead the values 0 and 1 of are obtained with a certain probability. 
The case of \autoref{eq:qubit}, $\ket{0}$ is measured with a probability of ${|\alpha|}^2$ and $\ket{1}$ with ${|\beta|}^2$.

\subsection{Depolarization Noise}
Since quantum computers are currently in the noisy intermediate state era (NISQ era)~\cite{preskill2018quantum}, quantum devices typically suffer from noise. We can describe this noise (specifically incoherent noise) by a linear and completely trace preserving map.
To make this more clear, let us consider that the pure quantum system is described by a density matrix $\rho$. 
Thus, we can describe the incoherent noise, so the noise from the environment by a linear and completely trace preserving map $\varepsilon(\rho)$. 
Hereby a convenient strategy is to employ the Kraus operators, so a set of operators ${K_i}$, with the property: $\sum_i K_i^\dagger K_i \leq I$~\footnote{Note that $A^\dagger$ describes the conjugate transpose of the matrix $A$.}. With those operators, the quantum channel can be described as:
\begin{equation}
    \varepsilon(\rho) = \sum_i K_i^\dagger \rho K_i.
\end{equation}
This concept can describe errors naturally occurring on the quantum device, like Phase-Flips, Bit-Flips or depolarization noise. Since in this work we will focus on the effect of depolarization noise in quantum channels, we will solely describe it here, omitting both phase and bit-flips. We refer the interested reader towards \citet{nielsen_chuang_2010} for further details.  

For a single qubit, the Kraus operators \(K_i\) representing the depolarizing channel can is given in terms of the Pauli matrices \(I\), \(X\), \(Y\), and \(Z\), and the depolarizing parameter \(p\). The parameter \(p\) usually denotes the probability that an error happens.
The Kraus operators for the depolarizing channel are:
\begin{align*}
    K_0 = \sqrt{1 - p} I & & K_1 = \sqrt{\frac{p}{3}} X \\
    K_2 = \sqrt{\frac{p}{3}} Y & & K_3 = \sqrt{\frac{p}{3}} Z
\end{align*}
Here, \(I\) is the identity operator and \(X\), \(Y\), and \(Z\) are the standard Pauli matrices:
\begin{align*}
    I = \begin{pmatrix} 1 & 0 \\ 0 & 1 \end{pmatrix} & & X = \begin{pmatrix} 0 & 1 \\ 1 & 0 \end{pmatrix} \\
    Y = \begin{pmatrix} 0 & -i \\ i & 0 \end{pmatrix} & & Z = \begin{pmatrix} 1 & 0 \\ 0 & -1 \end{pmatrix}
\end{align*}

The depolarizing noise is a type of quantum noise that affects qubits. It is a model for the process by which a qubit, with some probability, "forgets" its state and becomes a completely mixed state.

\subsection{Quantum Neural Network}

In the setting of quantum computing, QNNs are realized through what is known as a variational quantum algorithm (VQA)~\cite{cerezo2021variational}. The execution of these algorithms is a blend of quantum and classical processes. On the one hand, the learning model is described entirely by a quantum circuit or integrates a quantum circuit at a certain point. The optimization on the other hand is managed using classical methods. A central element in this quantum algorithmic structure is the parametrized quantum circuit (PQC). 
Composed of unitary transformations, for example including rotation gates:
\begin{align}
    R_z(\theta) &= \begin{pmatrix}
    e^{-i\theta/2} & 0 \\
    0 & e^{i\theta/2} \\
    \end{pmatrix}\label{eq:rz}, \\
    R_y(\theta) &= \begin{pmatrix}
    \cos(\theta/2) & -\sin(\theta/2) \\
    \sin(\theta/2) & \cos(\theta/2) \\
    \end{pmatrix}\label{eq:ry}, \\
    R_x(\theta) &= \begin{pmatrix} \cos(\frac{\theta}{2}) & -i\sin(\frac{\theta}{2}) \\ -i\sin(\frac{\theta}{2}) & \cos(\frac{\theta}{2}) \end{pmatrix}
\end{align}

these circuits can be adjusted via the parameters $\theta$ affecting the rotation gates. 
Regarding training, the loss function is computed using the true labels and the measurements of the quantum device. 

In our work we will assume a QNN, which consists out of three major parts. At first, we encode the individual datapoints $x \in \mathbb{R}^n$ onto the quantum circuit. Hereby a common choice is amplitude encoding using the Mottonen state encoding~\cite{mottonen2004transformation}. 
Amplitude encoding assumes the normalized input vector and provides a quantum circuit, which will encode the individual elements of $x$ onto the amplitudes of the quantum device: 
\begin{equation}
\ket{\psi_x} = \frac{1}{||x||} \sum_{i=1}^n x_i \ket{i}.
\end{equation}
This strategy is quite attractive, since it provides an exponential reduction of qubits for encoding our datapoints. 
As a next step a circuit ansatz with the classically optimizable parameters is provided. For this ansatz, numerous options exist~\cite{Huggins_2019,Cong_2019,wang2021exploration,Schuld_2020}. 
We select the ansatz of \citet{Schuld_2020} to be architecture-wise in line with \citet{Du_2021}.
The last part of the quantum neural network describes the measurement; the various classes typically are encoded on different basis states. Here, a set of qubits is selected to be measured. The measured outcomes a assigned to class probabilities. With those probabilities, the loss of the network can be computed and the parameters can be optimized by a classical optimizer. Note that a computation of the gradient is possible on a quantum device, by the so called "parameter-shift" rule.

\subsection{Quantum Adversarial Robustness}
Quantum adversarial robustness examines how specially crafted attacks can weaken a quantum-based machine learning model, just as traditional neural networks can be fooled by adversarial inputs~\cite{goodfellow2015explaining, kurakin2017adversarial, madry2019deep}. While there are other means of attacking a machine learning based system, we will focus on evasion attacks on the quantum machine learning model.
In the context of classical machine learning, adversarial robustness refers to the ability of a machine learning to maintain a stable prediction when confronted with adversarial examples. 

Formally, let $\mathcal{Y}$ denote the set of classes and $\|\cdot\|$ denote the euclidean norm on $\mathbb{R}^n$. Let $f:\mathbb{R}^n\rightarrow [0,1]^{|\mathcal{Y}|}$ be a \textit{classifier}, $x\in\mathbb{R}^n$ and $\epsilon>0$.
\smallskip

\begin{definition}[Adversarial Robustness]\label{def:adv_rob}
    Consider a classifier $f$ that maps an input $x$ to an output $y$. The adversarial robustness at input $x$ can be defined as the smallest amount of perturbation $\delta$ needed to change the model's prediction:
    \begin{equation*} 
    \operatorname{argmax}_{c\in\mathcal{Y}} f_c(x+\delta) \neq \operatorname{argmax}_{c\in\mathcal{Y}} f_c(x) \},
    \end{equation*}
    where $\|\delta\| \leq \epsilon$ is bounded by a maximum $\epsilon$ budget and ${x}_{adv}=x+\delta$ denotes an adversarial example.
\end{definition}
\smallskip

In the context of quantum computing, a study by \citet{lu2020quantum} found that QNNs can fall prey to these targeted perturbations. As part of their work, the quantum versions of the Fast Gradient Sign Method (FGSM)~\cite{goodfellow2015explaining} and Projected Gradient Descent (PGD)~\cite{aleks2017deep} were introduced.
Let us denote $x \in \mathbb{R}^n$ as (classical) features, $\mathcal{L}$ as loss function incorporating the QNN weights and $\nabla_x$ as gradient.
In the context of PGD, we define the number of iterations as $T$ and the adversarial budget as $\epsilon$.
Formally, we can obtain an adversarial example by iteratively computing:
$$
x_{adv}^{t+1} = \Pi_{x + \mathcal{S}} (x_{adv}^{t} + \frac{\epsilon}{T} \cdot \mathrm{sign}(\nabla_x \mathcal{L})),
$$
where $\Pi_{x + \mathcal{S}}(x')$ clips the value of $x'$ towards the space of adversarial perturbations around $x$. 
After $T$ iterations, the adversarial example can be found as $x_{adv}^{(T)}$.

FGSM on the other hand, assumes $T=1$ and hence computes once:
$$
x_{adv} = \Pi_{x + \mathcal{S}} (x + \epsilon \cdot \mathrm{sign}(\nabla_x \mathcal{L})),
$$
Typically the space of adversarial perturbations is defined by a Norm-Ball. This Norm-Ball on which PGD and FGSM attack the network is specified by selecting a specific process for $\Pi_{x + \mathcal{S}}(x')$. When using the sign function, the examples will act on the $L_\infty$-Norm. 
The fact that QNNs are vulnerable to adversarial perturbations motivates the discussion about specific measures to counteract adversarial examples. 
In this work, we will utilize \textit{adversarial accuracy} as in \autoref{def:adv_acc} to describe the network's vulnerability.

\begin{definition}[Adversarial Accuracy]\label{def:adv_acc}
   Let $\mathcal{D}=\{x_i,{y}_i\}_{i=1,\dots,N}$ be a dataset and $a(\cdot,f,\epsilon):\mathbb{R}^n\rightarrow\mathbb{R}^n$ be an adversarial model that tries to find an adversary within the perturbation budget $\epsilon$. Then, the \textit{adversarial accuracy} of the classifier $f$ on $\mathcal{D}$ under attack $a(\cdot,f,\epsilon)$ is defined as:
    \begin{equation}
        \frac{1}{N}\sum_{i=1}^N \mathbbm{1}({y}_i = f(a(x_i,f,\epsilon)))
    \end{equation}
\end{definition}
\smallskip

Where $\mathbbm{1}$ denotes the indicator function which corresponds to 1 if the argument is true, else is returning 0.
\subsection{Adversarial Defense Strategies}
Over recent years, multiple strategies have been developed to make models more robust against adversarial examples. In this context, we focused on two adversarial defense strategies:  adversarial training and randomized smoothing
\paragraph{Adversarial Training}
Adversarial training is a technique to improve the robustness of machine learning models against adversarial examples. It involves training the model on a mixture of regular and adversarially perturbed data, allowing the model to learn from and adapt to these perturbations. This method has been extensively studied, with seminal works by \citet{goodfellow2015explaining}, \citet{kurakin2017adversarial} and \citet{madry2019deep} being particularly influential in demonstrating its effectiveness and exploring its limits in various contexts.

\paragraph{Randomized Smoothing}
Randomized smoothing, not only a method for robustness certification but also an adversarial training technique~\cite{salman2020provably}, involves adding noise to training data. This noise, typically Gaussian or Uniform with zero mean, mathematically expressed as \( x = x + \delta \), with \( \delta\sim \mathcal{N}(0,\sigma^2I) \) or \( \delta\sim \mathcal{U}(-\sigma, \sigma) \), alters the training data. Unlike standard adversarial training that learns the worst-case decision boundary, randomized smoothing guides the classifier to learn a smoothed decision boundary. This approach integrates Gaussian perturbations with adversarial examples, not just enhancing empirical robustness but also significantly boosting certifiable robust accuracy of the smoothed classifiers.

\vspace{1em}
\begin{definition}[Certified Robustness]\label{def:rob}
     We say $f(x)\in\mathbb{R}^n$ is \emph{certifiable robust} with \emph{perturbation budget} $\epsilon > 0$ if
    \begin{equation}
        \operatorname{argmax}_{c\in\mathcal{Y}} f_c(x)=\operatorname{argmax}_{c\in\mathcal{Y}} f_c(x+\delta),
    \end{equation}
    holds for every $\delta\in\mathbb{R}^n\ \text{with}\ \|\delta\|<\epsilon$.
\end{definition}
\vspace{1em}

Differently from adversarial robustness (\autoref{def:adv_rob}), certified robustness has a stronger assurance: it provides a formal guarantee that a model's predictions will not change for any input within a specified range of perturbations.

\subsection{Quantum Certified Robustness}

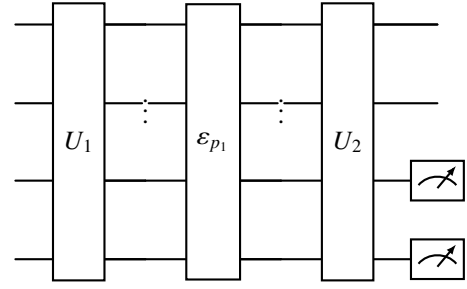
\begin{figure}[hbt]
    \centering
    \begin{quantikz}
        \qw     & \gate[4, nwires=2]{U_1} & \qw       & \gate[4, nwires=2]{\varepsilon_{p_1}} & \qw       & \gate[4, nwires=2]{U_2} & \qw \\
                &               & \vdots    &               & \vdots    &               &  \\
        \qw     &               & \qw       &               & \qw       &               & \meter{} \\
        \qw     &               & \qw       &               & \qw       &               & \meter{} \\
    \end{quantikz}
    \caption{Example Circuit with depolarization channel $\varepsilon_{p_1}$ and two final measurements.}
    \label{fig:examplecircuit}
    \vspace{-2em}
\end{figure}

In the context of QML, the robustness bound is motivated by so called quantum differential privacy. Similarly to classical differential privacy, quantum differential privacy assumes two datapoints $\sigma$ and $\rho$, separated by a small \textit{trace distance}: $\tau(\sigma, \rho) = \sqrt{1-\text{Tr}(\sigma, \rho)}$ and for all measurable sets $S$ it must hold that:
\begin{equation}\label{eq:differential_privacy}
    \text{Pr}\left( M(\sigma) \in S \right) \leq e^{\epsilon} \text{Pr}\left( M(\rho) \in S \right) +\delta
\end{equation}
Where $(\epsilon, \delta)$ is referred to as the privacy budget of the algorithm. $M(x)$ is the quantum classifier in our context. Intuitively, differential privacy in this sense describes that the output of the algorithm differs only slightly due to a small change of input information.
In this context, \citet{Du_2021} have provided a robustness bound for quantum classifiers based on depolarization noise by assuming two quantum states, $\sigma$, and $\rho$, which are the encoded data. The \textit{trace-distance}, between those two states is then upper-bounded by $\tau_D$: $\tau(\sigma, \rho) \leq \tau_D$.
Note that throughout this work, we assume the infinite sampling case of their upper bound since all models are based on state-vector simulation rather than actual sampling of a quantum simulator or physical device. Next, assume that we are provided with the class probability $y_C(\sigma)$ of our class $C$~\footnote{Assume $C=0$ without loss of generality.}, which we can obtain by measuring the final circuit. We define the follow-up class $y_{k \neq C}(\sigma) = \argmax_{k \neq C} \left\{ y_k(\sigma) \right\}$ and can relate those two classes via quantum differential privacy by \autoref{eq:yrelation}. 
\begin{align}
    y_C(\sigma) > e^{2\epsilon} y_{k \neq C}(\sigma) \label{eq:yrelation} \\
    \epsilon = \ln \left( 1 + D_{meas}\nicefrac{(1-p) \tau_D}{p}\right) \label{eq:eps_definition}
\end{align}
Where $\epsilon$ is defined as \autoref{eq:eps_definition}. In this term, $p = 1 - \prod_{i=1}^j (1-p_i)$ is the multiplication of all probabilities of depolarization noise occurrences in each depolarization channel (refer to \autoref{fig:examplecircuit} for an example) and  $D_{meas} \geq K$ serves as an upper bound of the dimensionality $K$ of the measurement operator. To clarify this, assume we are provided with $2$ measurements as in \autoref{fig:examplecircuit}, our measurement operator would have a dimensionality of $2^2 = 4$. 

\section{Robustness Analysis}
As mentioned earlier, our goal is to analyze the robustness of QNNs and the effect of the bound provided by \citet{Du_2021}. Consequently, we examined the following three research questions, which we deemed relevant:
\begin{enumerate}
    \item What is the effect of depolarizing noise on the adversarial accuracy?
    \item How does this effect compare to state-of-the-art defense strategies in our setting?
    \item Will a blend of depolarizing noise and state-of-the-art defense strategies, have an effect?
\end{enumerate}
To answer those questions, we utilized the dataset used by \citet{Du_2021} to outline the reproducability of his results. Secondly we used the MNIST~\cite{LeCun1998}, Breast Cancer Wisconsin (BC)~\cite{bischl2021openml} and Pima Indians Diabetes Database(PID)~\cite{bischl2021openml}. We wanted to estimated the performance of CIFAR10, by truancing it towards classes zero and nine. We train all models on the ansatz proposed by ~\citet{Schuld_2020} and encode our data on the quantum circuit using amplitude encoding. 

\subsection{Data-Preprocessing}
We will outline the data preprocessing done for all the datasets. Typically this process included dataset specific preprocessing techniques (normalization, feature engineering) combined with a split into a train, validation and test set or into a train and test set, depending on the size of the dataset. We performed this split the partition was varied among the datasets~\footnote{Splitting of the datasets was done at random, utilizing sklearn's \texttt{train\_test\_split} function. For multiple splits we separated the validation and test set from the validation set.}

\subsubsection{Iris}
For Iris, we followed the pre-processing of \citet{Du_2021}, which resulted in the removal of the \textit{versicolor} species and the \textit{petal\_width} feature. For the resulting dataset, we reported 100 feature vectors $x \in \mathbb{R}^3$, which were padded by a $0$ and normalized by their respective $L_2$-Norm. Splitting the dataset into a train and test set resulted in 60 training samples and 40 test samples. As a last step, we one-hot encoded class labels to make them compatible with the resulting measurement outcome from the quantum device.

\subsubsection{Pima Indians Diabetes (PID)}
For the Pima Indians Diabetes dataset, we first dropped all nan or duplicate features from the dataset, next to overcome the present class imbalance, we removed overall $232$ datapoints from the majority class at random. We used the remaining dataset with $536$ datapoints and split them into a train set consisting of $375$ datapoints and a test set of $161$ datapoints. The train and test set was standard normalized ($x = \nicefrac{x - x_{\text{mean}}}{x_{\text{std}}}$)\footnote{We used the \texttt{StandardScaler} from the sklearn package} and afterwards rescaled to the interval $\left(0;1\right]$ to allow a comparison among other datasets. 
\subsubsection{Breast Cancer Wisconsin (BC)}
For the Breast Cancer Wisconsin dataset, we dropped the nan features and duplicates as for the Pima Indians Dataset. This resulted in overall $683$ features. During hyperparameter training of the QNN, we figured that standard normalization hindered the convergence of the model, which was the reason why after splitting the dataset into a train set ($546$ datapoints) and test set ($137$ datapoints) we only rescaled the data into the interval $\left(0;1\right]$. It was not necessary to drop datapoints of the majority class, since the BC dataset was quite balanced from our point of view (we found a ratio of $65:35$, which we considered acceptable given the $\approx 90\%$ accuracy on the test set)
\subsubsection{MNIST}
For the MNIST10 dataset, we rescaled the $28\times28$ images into smaller images of the size $16\times16$. Flattening the image resulted in a feature vector $x \in \mathbb{R}^{256}$, which can be realized with eight qubits. Next, we normalized the individual elements of the feature vector to be in the range $[0;1]$ and finally split the data into a training set of $48.000$ images, a validation set of $10.800$ images, and a test set of $1200$ images. 
To provide additional insight, specific classes of the dataset were subselected to mimic a binary classification task, MNIST2, and a four-class multiclassification task, MNIST4. For MNIST2, we used the classes for the digits one and nine. For MNIST4, we extended the present labels in MNIST2, providing overall the labels one, nine, seven, and three.
MNIST2 consisted of $12.691$ datapoints, split into $10.152$ training images, $2285$ validation images, and $254$ testing images. Analougus MNIST4, consisting of $25.087$ images, was split into train-, test- and validation-set with the respective sizes $20.069$, $4516$, and $502$.

\subsubsection{CIFAR2}
For CIFAR10~\cite{krizhevsky2009learning}, we droped all classes except zero and nine, upscaled the images from size $32\times32$ to $64\times64$, after this rescaling, we computed the histogram of orieanted gradients (HOG) using the scikit-image library~\cite{scikit_image}. For computing the HOG-features we used a $8\times8$ cells with a inner block size of $1\times1$. Overall this resulted in a feature vector $x \in \mathbb{R}^{576}$. We split the dataset into a train set of size $8000$ a validation set of size $1800$ and a test set of size $200$. We labeled this dataset CIFAR2, since we converted the multiclass classification task to a binary classification task. 

\subsection{Model Architecture}

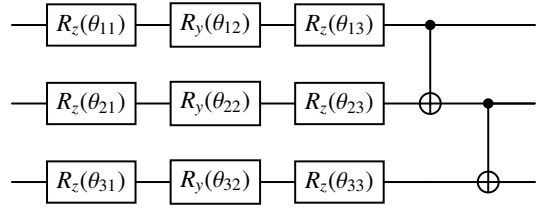
\begin{figure}
    \centering
    \begin{adjustbox}{width=0.8\linewidth}
            \begin{quantikz}
                \qw & \gate{R_z(\theta_{11})} & \gate{R_y(\theta_{12})} & \gate{R_z(\theta_{13})} & \ctrl{1}  & \qw       & \qw \\
                \qw & \gate{R_z(\theta_{21})} & \gate{R_y(\theta_{22})} & \gate{R_z(\theta_{23})} & \targ{}   & \ctrl{1}  & \qw \\
                \qw & \gate{R_z(\theta_{31})} & \gate{R_y(\theta_{32})} & \gate{R_z(\theta_{33})} & \qw       & \targ{}   & \qw \\
            \end{quantikz}
    \end{adjustbox}
    \caption{Strongly Entangling Layer architecture by \citet{Schuld_2020}. In this quantum circuit outlined for three qubits. Note that the control and target qubits of the entangling layer may vary up to a permutation by different layers.}
    \label{fig:strong_ent_layers}
    \vspace{-1em}
\end{figure}

As pointed out earlier, we used the ansatz of \citet{Schuld_2020}, which consists of \textit{strongly entangling layers} outlined in \autoref{fig:strong_ent_layers}. Those layers typically consist of three rotation gates, in our case, a combination of $R_z$ and $R_y$ rotations and a sequence of CNOT-Gates, which are permuted differently on the wires for each layer. The reason for our choice of this particular ansatz is that it provides a simple yet adaptive structure. The number of required qubits and measurements can be determined from the input data size. In contrast, the number of layers influences the amount of parameters and the model's capacity. One can find the hyper-parameters used in this study in \autoref{tab:hyperparams}. The Table also shows the resulting circuit depth, as well as the single- and two-qubit gate count for each circuit, resulting from the selected hyperparameters.

\begin{table*}[hbt]
    \centering
    \caption{Hyperparameters used for the various networks for the different datasets. We used the \texttt{qiskit.aer} device and pennylanes specs. Single Qubit gates are $R_z$ and $R_y$, two qubit gates are $CX$. }\label{tab:hyperparams}
    \begin{adjustbox}{width=0.8\linewidth}
        \begin{tabular}{lrrrrrrr}
            \toprule
            Hyper-Parameter             & Iris & PID & BC  & MNIST2    & MNIST4 & MNIST10  & CIFAR2 \\ 
            \midrule
            learning rate               & 0.05 & 0.0005 & 0.0005  & 0.005     & 0.005 & 0.005     & 0.0005  \\
            batch size                  & 30   & 16     & 16      & 512       & 512   & 128       & 64 \\
            Nr. epochs                  & 100  & 10     & 10      & 30        & 30    & 30        & 10 \\
            Nr layers                   & 2    & 40     & 40      & 40        & 40    & 40        & 80\\
            Nr. measurements            & 1    & 1      & 1       & 1         & 2     & 4         & 1\\
            \midrule \midrule
            Nr. Parameters              & 13   & 181 & 481 & 961 & 961 & 961  & 2641\\
            Circuit Depth               & 11    & 121 & 242 & 276 & 276 & 276  & 601\\
            Single Qubit Gates          & 12   & 180 & 480 & 960 & 960 & 960  & 2640\\
            Two Qubit Gates             & 5    & 60 & 160 & 320 & 320 & 320  & 880\\
            \bottomrule
        \end{tabular}
    \end{adjustbox}
\end{table*}

We want to re-emphasize that specifically for MNIST10, a different architecture for an ansatz specifically exists, which achieved a $~90\%$ test accuracy~\cite{wang2021exploration}. Nevertheless, we wanted to provide a coherent architecture throughout this work to reasonably compare the QNNs among datasets. 

Additionally, due to the noisy architecture and limited availability of quantum devices, we have simulated the quantum device using statevector simulation. Specifically, we utilized the auto-grad feature of pytorch and the pennylane library to formulate the ansatz~\footnote{Pennylane~\cite{bergholm2022pennylane} 0.29.1 and pytorch~\cite{NEURIPS2019_9015} 2.0.0}.

\subsection{Robustness Upperbound}
During our analysis, we want to compare the bound obtained by \citet{Du_2021} with empirical results found in our study. 
Therefore, we utilize \autoref{eq:yrelation} and rewrite it towards obtaining a lower bound for $\epsilon$:
\begin{align*}
    \nicefrac{y_C(\sigma)}{y_{C \neq k}(\sigma)} < e^{2\epsilon} \\
    \frac{1}{2}\ln \left( \nicefrac{y_C(\sigma)}{y_{C \neq k}(\sigma)} \right) < \epsilon
\end{align*}
Hence, this lower bound for $\epsilon$ is given as follows:
\begin{equation}\label{eq:eminlowerbound}
    \epsilon_{min} = \frac{1}{2}\ln \left( \nicefrac{y_C(\sigma)}{y_{C \neq k}(\sigma)} \right)
\end{equation}
Combining \autoref{eq:eminlowerbound} and \autoref{eq:eps_definition}, we can obtain the following:
\begin{equation}\label{eq:epsilon_min}
    \epsilon_{min} = \ln \left( 1 + D_{meas}\nicefrac{(1-p) \tau_D}{p}\right)
\end{equation}
We can rewrite this to obtain the certified \textit{trace distance} in terms of the quantum states as follows:
\begin{align*}
    e^{\epsilon_{min}} = 1 + D_{meas}\nicefrac{(1-p) \tau_D}{p}, \\
    \frac{e^{\epsilon_{min}}-1}{D_{meas}} \cdot \frac{p}{(1-p)} = \tau_D \\
\end{align*}
Hence, we can obtain the certified distance $\tau_D$ as follows:
\begin{equation}
    \tau_D = \frac{e^{\epsilon_{min}}-1}{D_{meas}} \cdot \frac{p}{(1-p)}
\end{equation}
Assuming that the probability of depolarization noise $p$ is known, we can see that the certified distance $\tau_D$ is only dependent on the fraction of the class probability and to the follow-up class, the size of the measurement operator.

In our study, we will attack the quantum classifier with the FGSM and PGD attacks\footnote{FGSM for the MNIST-Classifiers and PGD for the Iris Classifier. We utilized the FGSM attack since computing the gradient of the quantum circuit concerning the input is quite resource-intensive when a noise channel is evaluated.}, both bounded by the $L_\infty$-norm. Hence, we need to ensure that the certified distance we will obtain from our classification is an upper bound to those two norms. \citet{Du_2021} have already argued, that the $L_2$-Norm is bounded by the trace distance:
\begin{equation}
    L_2 \leq \tau(\sigma, \rho)
\end{equation}
Consequently, we can utilize $L_\infty \leq L_2$ to argue that this bound is also valid for attacks using an $L_\infty$-Norm. 
We did not normalize the data points in the MNIST datasets by the $L_2$-norm but performed normalization during the encoding. A comparison of this bound towards the classical epsilon value is hence difficult, since attacks will be made in the unnormalized space. Nevertheless, the upper bound can provide us with an estimate of the certifiable robustness of the data points and a certain degree of explanatory power. To make this more clear, assume a correct classification, which will yield the relation: $y_C(\sigma) > y_{C \neq k}(\sigma)$. We can conclude from this relation, that the fraction $\nicefrac{y_C(\sigma)}{y_{C \neq k}(\sigma)}$ will be greater than one: $\nicefrac{y_C(\sigma)}{y_{C \neq k}(\sigma)} > 1 $. Hence, the natural logarithm in \autoref{eq:epsilon_min} will also be greater or equal to zero. Since $e^0 = 1$, approaching $e^{\epsilon_{min}}-1$ will result in a positive upper bound.
On the other hand, in case we are provided with a misclassification, the class probability will be smaller than the follow-up class: $y_C(\sigma) < y_{C \neq k}(\sigma)$. Hence, the fraction $\nicefrac{y_C(\sigma)}{y_{C \neq k}(\sigma)}$ will be smaller than one, and the natural logarithm will be negative. This will result in $e^{\epsilon_{min}}-1 < 0$; so, the certifiable distance will be smaller than $0$. We interpret this as it is impossible to certify the data point based on the model's output.

Note that this reasoning indicates that a model with more parameters and good generalization can be more robust. This may be achieved by quantum convolutional neural networks or a different design of a variational ansatz.

\section{Experiments}
With our experiments, we try to answer the three questions raised above. At first, we want to see the influence of depolarization noise on a multi-class classifier, therefore, we used the Iris, MNIST2, MNIST4 and MNIST10 datasets as well as the PID, BC and CIFAR2 datasets. Secondly, our interest shifts to whether one can defend our QNNs with other adversarial training methods, specifically randomized smoothing and adversarial training (which we refer to as PGD for reasons of conciseness) we hereby focused on the Iris, Breast Cancer, Pima Indians Diabetes, CIFAR2, MNIST2, MNIST4 and MNIST10 datasets. Finally, we want to elaborate on whether we can chain adversarial training with depolarization noise as a blend of two adversarial defense mechanisms, for this particular question we solely evaluated MNIST2.

\begin{table}
    \centering
    \caption{Test accuracy ($Acc_{test}$) for the different models trained during this study. For the iris dataset we reused the test as validation set during training due to the limited available datapoints. Additionally the mean certified distance is reported for the test set, with respect to different depolarization noise values: $[0.1, 0.5, 0.9]$. We marked the best mean certified distance per experiment in bold and the negative certified distances as red.}\label{tab:model_accuracy}
    \resizebox{\linewidth}{!}{%
    \begin{tabular}{ll | l | r | rrr}
        \toprule
        \textbf{Dataset}                         &\textbf{Train}                           & \textbf{Param.} & \textbf{Acc.}   & \multicolumn{3}{c}{\textbf{Certified Distance}} \\
        & & & & \multicolumn{3}{c}{$\tau_D$ ($\times 10^{-2}$)} \\
                                        &                                   &$\epsilon$ or $\sigma$           &            &  $p=0.1$  &  $p=0.5$  & $p=0.9$  \\
        \midrule
        PID           & vanilla                           & --        & 67.08           & \textbf{0.449} & 0.443 & 0.055 \\
        \midrule
        BC                   & vanilla                           & --        & 91.24         & \textbf{6.210} & 5.189 & 0.926 \\
        \midrule
       
        \multirow{7}{*}{Iris}           & vanilla                           & --        & 100.0         & 4.176 & \textbf{5.960} & 1.092 \\
        \cmidrule{2-7}
                                        &\multirow{3}{*}{PGD}      
                                                                            & 0.01      & 100.0         & 2.823 & 5.215 & 1.291 \\
                                        &                                   & 0.05      & 100.0         & 2.811 & 5.184 & 1.276 \\
                                        &                                   & 0.09      & 100.0         & 2.774 & 5.100 & 1.254 \\
        \cmidrule{2-7}
                                        &\multirow{3}{*}{RS}                & 0.1       & 100.0         & 2.824 & 5.181 & 1.273  \\
                                        &                                   & 0.5       & 97.50         & 2.661 & 4.412 & 1.029 \\
                                        &                                   & 0.9       & 70.00         & 2.375 & 3.502 & 0.723 \\
        \midrule
        \multirow{7}{*}{MNIST2}         & vanilla                           & --        & 99.61         & \textbf{2.429} & 0.150 & 0.006 \\
        \cmidrule{2-7}
                                        &\multirow{3}{*}{PGD}       
                                                                            & 0.01      & 99.61         & 2.366 & 0.135 & 0.003 \\
                                        &                                   & 0.05      & 99.61         & 2.372 & 0.127 & 0.001 \\
                                        &                                   & 0.09      & 99.21         & 2.392 & 0.123 & 0.002 \\
        \cmidrule{2-7}
                                        &\multirow{3}{*}{RS}                & 0.1       & 99.61         & 2.369 & 0.124 & 0.001 \\
                                        &                                   & 0.5       & 99.61         & 2.378 & 0.116 & {\color{red}-0.003} \\
                                        &                                   & 0.9       & 99.61         & 2.253 & 0.111 & 0.002 \\
        \midrule
        \multirow{7}{*}{MNIST4}         & vanilla                           & --        & 90.44         & \textbf{1.035} & 0.042 & {\color{red} -0.007} \\
        \cmidrule{2-7}
                                        &\multirow{3}{*}{PGD}       
                                                                            & 0.01      & 90.64         & 0.997 & 0.034 & {\color{red} -0.006} \\
                                        &                                   & 0.05      & 90.64         & 0.995 & 0.044 & {\color{red} -0.003}  \\
                                        &                                   & 0.09      & 90.24         & 1.000 & 0.037 & {\color{red} -0.005}  \\
        \cmidrule{2-7}
                                        &\multirow{3}{*}{RS}                & 0.1       & 91.24         & 0.988 & 0.048 & {\color{red} -0.004} \\
                                        &                                   & 0.5       & 91.04         & 0.988 & 0.047 & {\color{red} -0.003} \\
                                        &                                   & 0.9       & 91.43         & 0.916 & 0.050 & {\color{red} -0.003} \\
        \midrule
        \multirow{7}{*}{MNIST10}        & vanilla                           & --        & 80.00         & \textbf{0.205} & {\color{red} -0.012} & {\color{red} -0.007} \\
        \cmidrule{2-7}
                                        &\multirow{3}{*}{PGD}       
                                                                            & 0.01      & 80.50         & 0.172 & {\color{red} -0.049} & {\color{red} -0.014} \\
                                        &                                   & 0.05      & 80.08         & 0.173 & {\color{red} -0.049} & {\color{red} -0.015} \\
                                        &                                   & 0.09      & 81.75         & 0.164 & {\color{red} -0.030} & {\color{red} -0.011} \\
        \cmidrule{2-7}
                                        &\multirow{3}{*}{RS}                & 0.1       & 79.50         & 0.180 & {\color{red} -0.062} & {\color{red} -0.017} \\
                                        &                                   & 0.5       & 76.17         & 0.174 & {\color{red} -0.053} & {\color{red} -0.014} \\
                                        &                                   & 0.9       & 71.91         & 0.167 & {\color{red} -0.013} & {\color{red} -0.006} \\
        \midrule
        CIFAR2                          & vanilla                           & --        & 82.50         & \textbf{0.249} & 0.005 & 0.001 \\
        \bottomrule
    \end{tabular}
    }
\end{table}
\subsection{Training of the Models}

As already pointed out, our models all follow a \texttt{AmplitudeEmbedding} plus a \texttt{StronglyEntangling} layer structure and only vary in terms of the datasets. We train all models using Adam as an optimizer by varying the batch size, learning rate, and number of epochs. These models are referred to as vanilla models. PyTorch's categorical-cross entropy was utilized as the loss function for Iris, BC, PID, CIFAR2, MNIST2, and MNIST4. For the MNIST10 dataset, however, the application of negative log likelihood as the loss function, expressed as \(l(y, \hat{y}) = -\sum_i y_i \log (\hat{y}_i)\), was found to enhance accuracy. This effectiveness in the specific case of MNIST10 led to its selection as the preferred loss function for this dataset.
Regarding adversarial training, we train the models using the PGD attack with \(L_\infty\)-norm for the Iris dataset with \(T=50\) steps. Due to performance issues, we train the MNIST2, MNIST4, and MNIST10 models with \(T=3\) steps. 
We also employ randomized smoothing during training; thus, we perturb the original image with noise from a uniform distribution and use this as a training sample. The variance of this training is reported as \(\sigma\), and the mean is set to zero.
We employ early stopping for both randomized smoothing and adversarial training. We use the loss value at which the models are considered to be converged. For Iris, this value is set to \(0.6\). For MNIST2, the value is set to \(0.45\), for MNIST4 to \(1.15\), and for MNIST10 to \(1.7\). Note that we do not perform adversarial training on BC, PID, or CIFAR2, since these datasets are solely used to assess performance under depolarization noise.
The test accuracy, combined with the mean certified distance \(\tau_D\) derived from the test set, is reported in \autoref{tab:model_accuracy}.

\begin{figure*}[]
    \centering
    \begin{subfigure}{0.3\linewidth}
    \includegraphics[width=\textwidth]{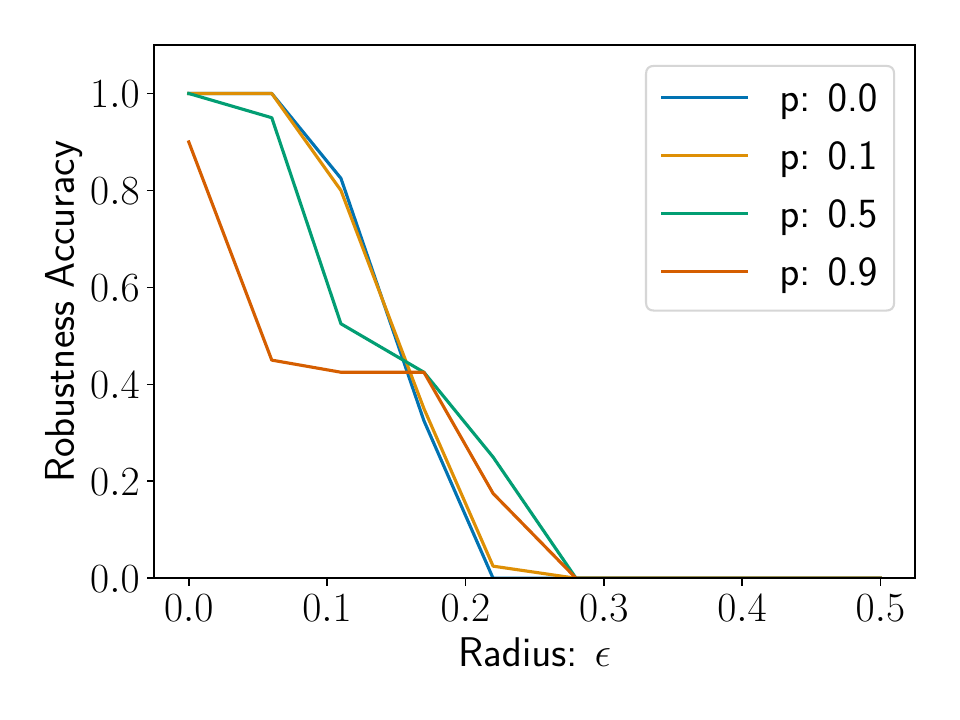}
    \caption{Iris}
    \end{subfigure}
    \begin{subfigure}{0.3\linewidth}
    \includegraphics[width=\textwidth]{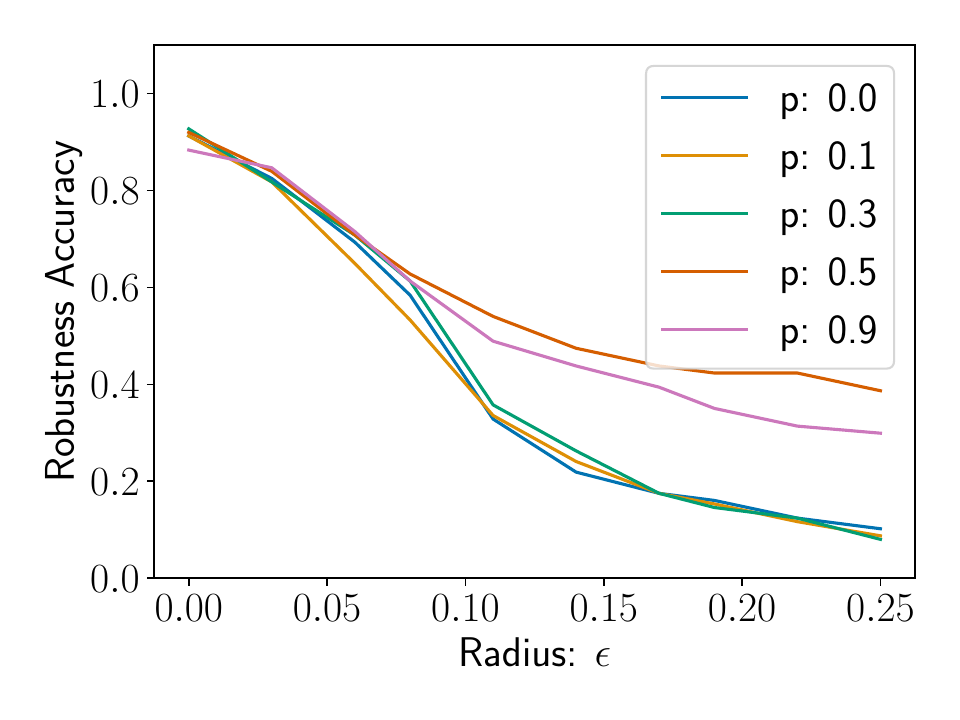}
    \caption{BC}
    \end{subfigure}
    \begin{subfigure}{0.3\linewidth}
    \includegraphics[width=\textwidth]{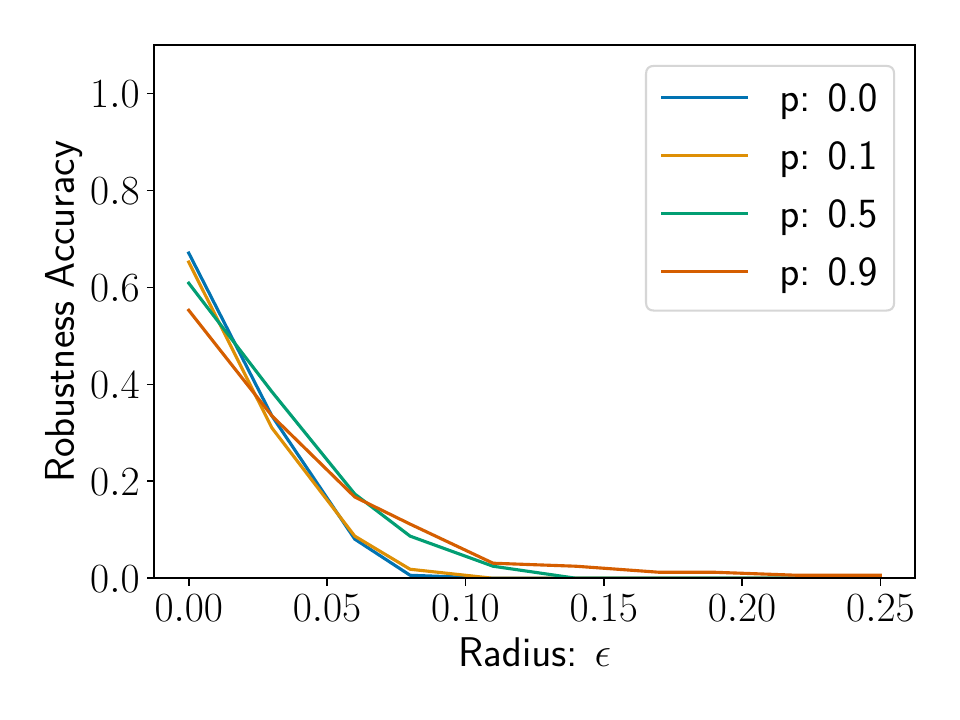}
    \caption{PID}
    \end{subfigure}
    \\
    \begin{subfigure}{0.3\linewidth}
    \includegraphics[width=\textwidth]{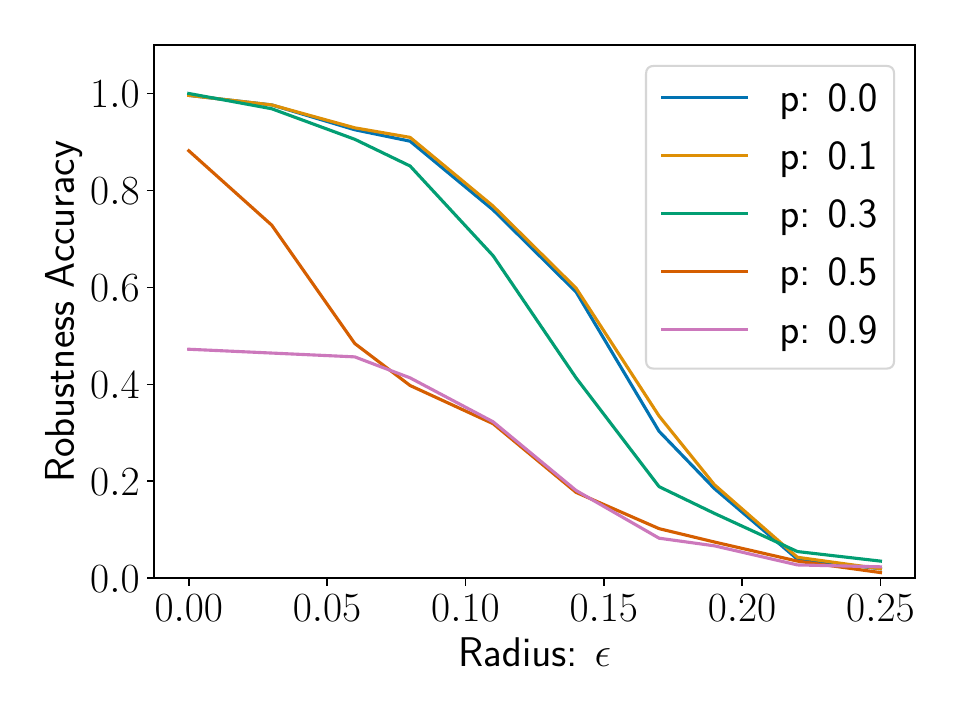}
    \caption{MNIST2}
    \end{subfigure}
    \begin{subfigure}{0.3\linewidth}
    \includegraphics[width=\textwidth]{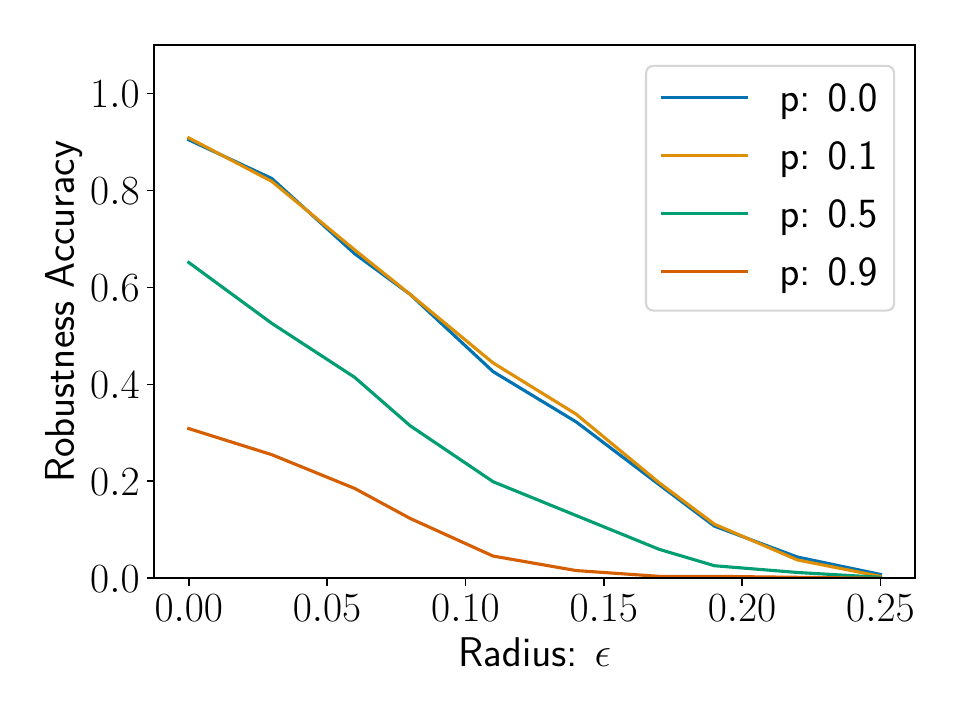}
    \caption{MNIST4}
    \end{subfigure}
    \begin{subfigure}{0.3\linewidth}
    \includegraphics[width=\textwidth]{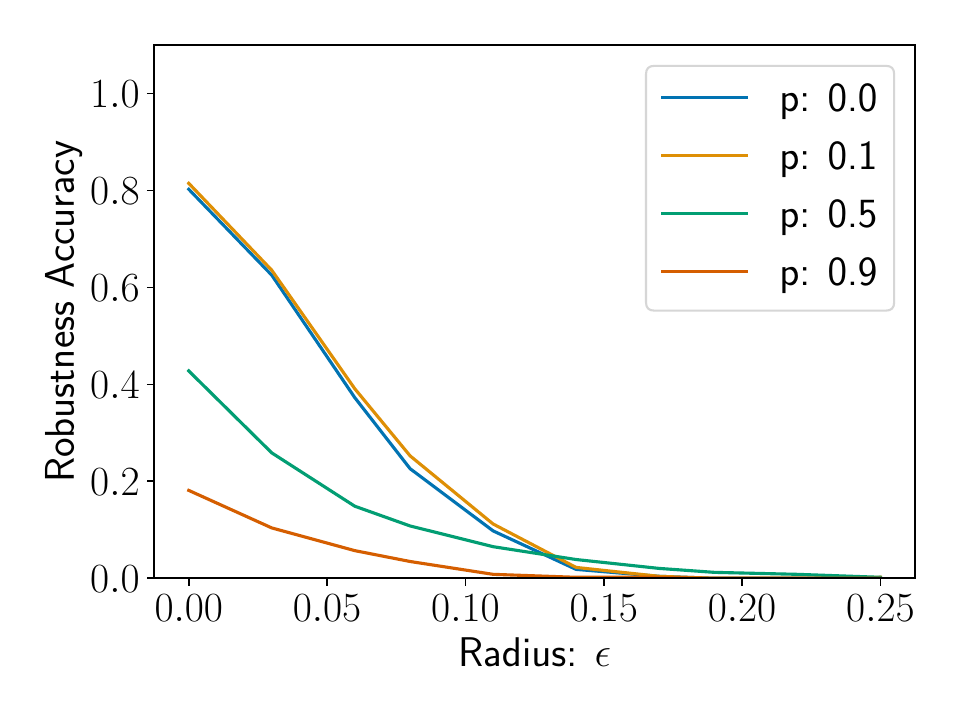}
    \caption{MNIST10}
    \end{subfigure}
    \vspace{0.5em}
    \caption{Robustness accuracy for the various vanilla classifiers for Iris, BC, PID, MNIST2, MNIST4 and MNIST10.}
    \label{fig:rq1_classifiercomparison}
\end{figure*}
\subsection{Effect of Depolarization Noise}

We compare the vanilla networks of the Iris, PID, BC, MNIST2, MNIST4, and MNIST10 datasets for depolarization noises \(p\) in the range \([0.1, 0.5, 0.9]\). We include a depolarizing noise of \(0.3\) for MNIST2.
Our results are shown in \autoref{fig:rq1_classifiercomparison}. 
Additionally, we report the adversarial accuracy of CIFAR2 for the depolarization factors of \([0.1, 0.5, 0.9]\) for \(\epsilon \in [0.0, 0.3]\). We choose the factor of \(0.3 \approx \nicefrac{8}{255}\), as this is also the factor used in the benchmarks conducted on CIFAR10. However, we note that our version of CIFAR2 is not comparable with the benchmarks done on CIFAR10, since we have converted the multi-class classification into a binary classification and our attack is conducted on the HOG features rather than on the actual pixels.

Regarding the binary classification networks BC, PID, Iris, and MNIST2, we typically observe that the QNN performs well in terms of accuracy. With PID, this is noteworthy as the QNN achieves only an accuracy of \(67.08\%\). Comparing this score to the random forest approach in \citet{pima_ref}, which resulted in an accuracy of \(79.57\%\), we see that this accuracy corresponds to the upper quartile of the experiments, as shown in \citet[][Fig. 1]{egginger2023hyperparameter}.

For Iris, BC, and PID, we observe a similar behavior as seen in \citet[][Apdx. H]{Du_2021}: Additional depolarization noise typically aids the classifier for larger epsilon values, although too much noise might decrease the initial accuracy. PID is particularly interesting, as it shows this effect even for small adversarial perturbations. However, the practical use of this effect is questionable, as it typically occurs when the classifier is already below or close to \(50\%\) accuracy.

In contrast, for MNIST2, we do not find any significant improvement in terms of adversarial accuracy. A similar behavior is observed for CIFAR2, as detailed in table \autoref{tab:cifar2}. For MNIST4 and MNIST10, we note that depolarization noise does not increase robustness. This is attributed to two factors.

First, the limited capacity of the network to learn the data distribution. This is supported by \autoref{tab:model_accuracy}, showing that accuracy decreases with an increase in the number of classes. Additionally, when examining the certified bounds, we notice that the minimum of all computed bounds is typically negative, and the maximum trace distance is usually smaller than for binary classification. The negative bound is attributed to our inability to achieve \(100\%\) accuracy on the test set.

Secondly, for MNIST4 and MNIST10, we measure two qubits to obtain the classes for MNIST4 and four qubits for MNIST10. The poor performance could also be attributed to depolarization noise automatically increasing the probability of the follow-up class, since two classes are assigned to one qubit. However, this assumption requires further experimental and analytical evaluation, which is a subject for future work.

\begin{table}[]
    \centering
    \caption{Mean Robustness Accuracy for CIFAR2 for various depolarization noises}\label{tab:cifar2}
    \begin{tabular}{l|r r}
    \toprule
        Depol. & \multicolumn{2}{c}{Robustness Accuracy (\%)}  \\
        $p$ &$\epsilon=0.0$ & $\epsilon=0.03$ \\
        \midrule
         0.0 & 82.5 & 56.0 \\ 
         0.1 & 79.0  & 58.0 \\
         0.5 & 49.0  & 49.0 \\
         0.9 & 49.0  & 49.0 \\
         \bottomrule
    \end{tabular}
\end{table}

\subsection{Effect of Adversarial Training}
\begin{figure}[!hbt]
    \centering
    \begin{subfigure}{0.49\linewidth}
        \includegraphics[width=\textwidth]{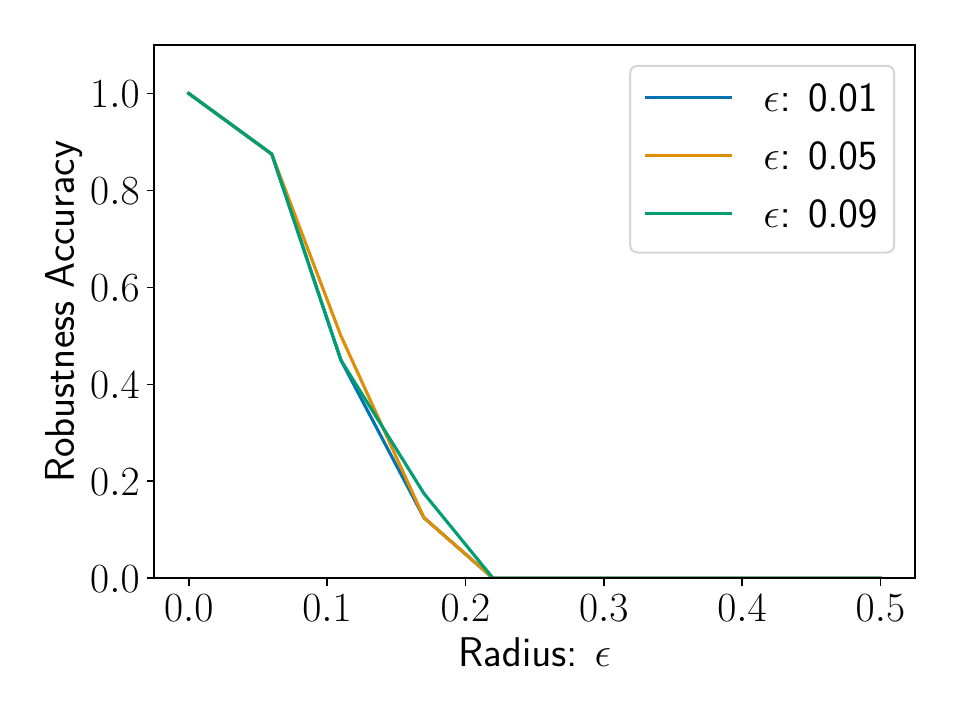}
        \caption{Iris: PGD}
    \end{subfigure}
    \begin{subfigure}{0.49\linewidth}
        \includegraphics[width=\textwidth]{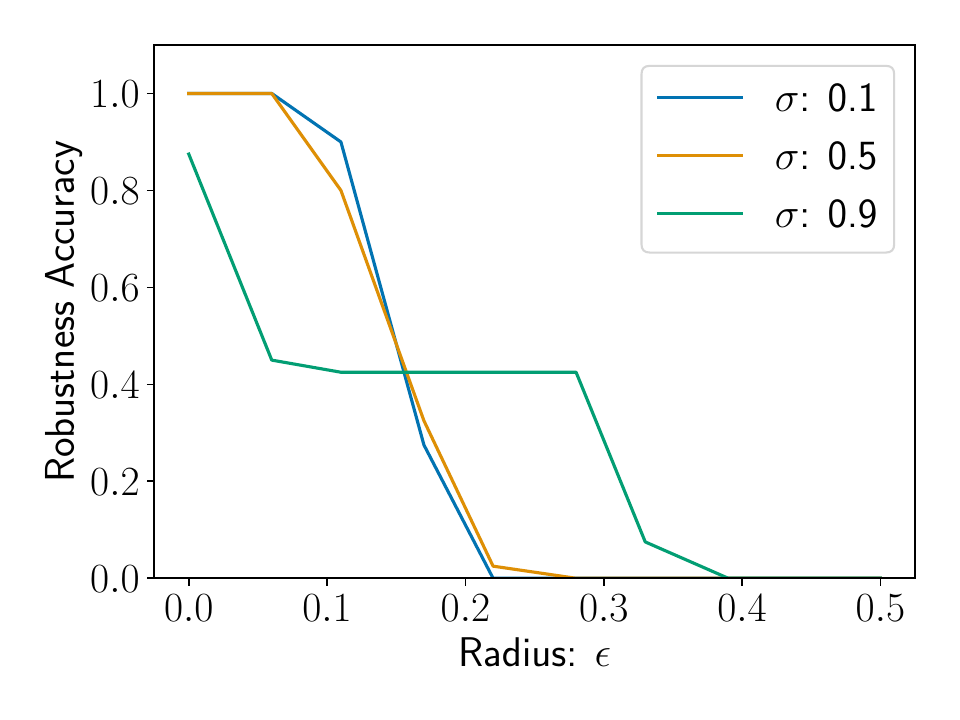}
        \caption{Iris: RS}
    \end{subfigure}
    \caption{Comparison of the two methods randomized smoothing (RS) and adversarial training (PGD) specifically for the Iris dataset.}
    \label{fig:iris_adversarial}
\end{figure}

\begin{figure*}
    \centering
    \begin{subfigure}{0.3\linewidth}
        \includegraphics[width=\textwidth]{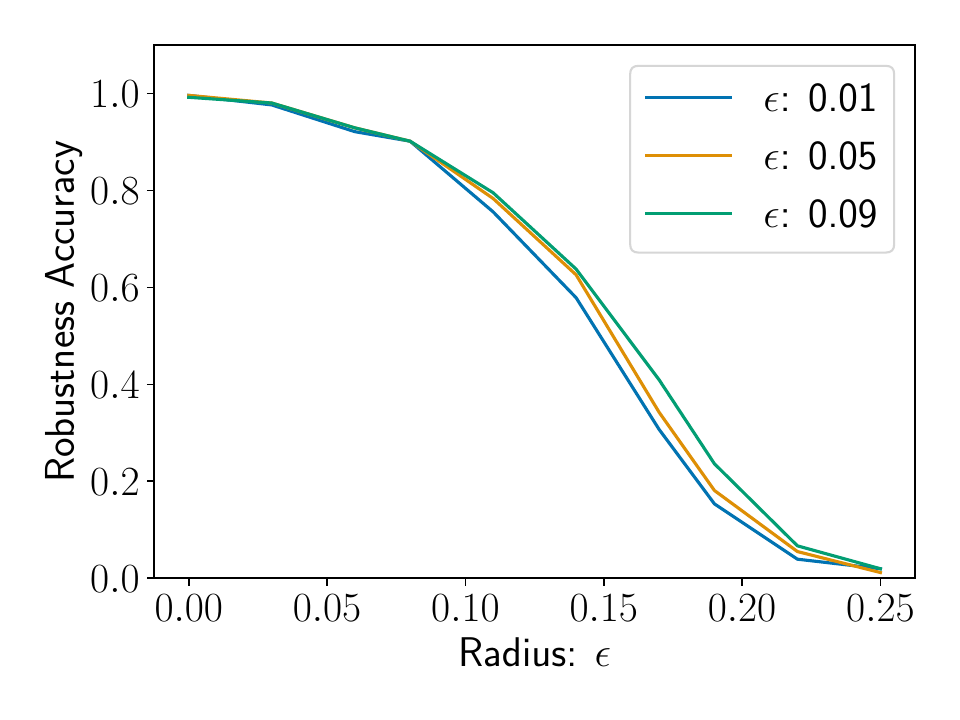}
        \caption{MNIST2: PGD}
    \end{subfigure}
    \begin{subfigure}{0.3\linewidth}
        \includegraphics[width=\textwidth]{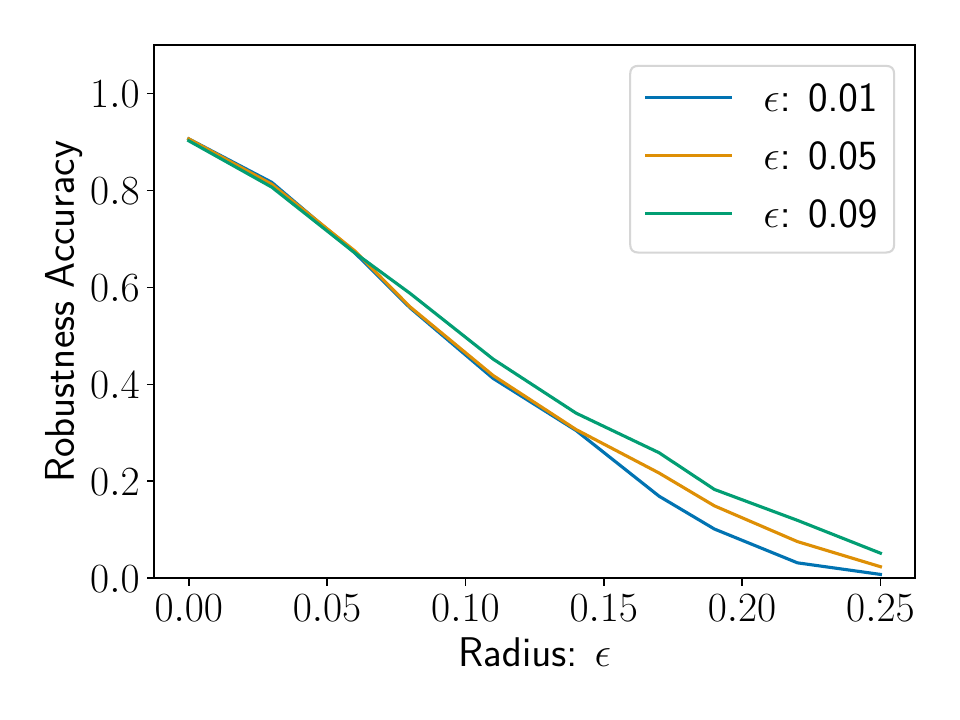}
        \caption{MNIST4: PGD}
    \end{subfigure}
    \begin{subfigure}{0.3\linewidth}
        \includegraphics[width=\textwidth]{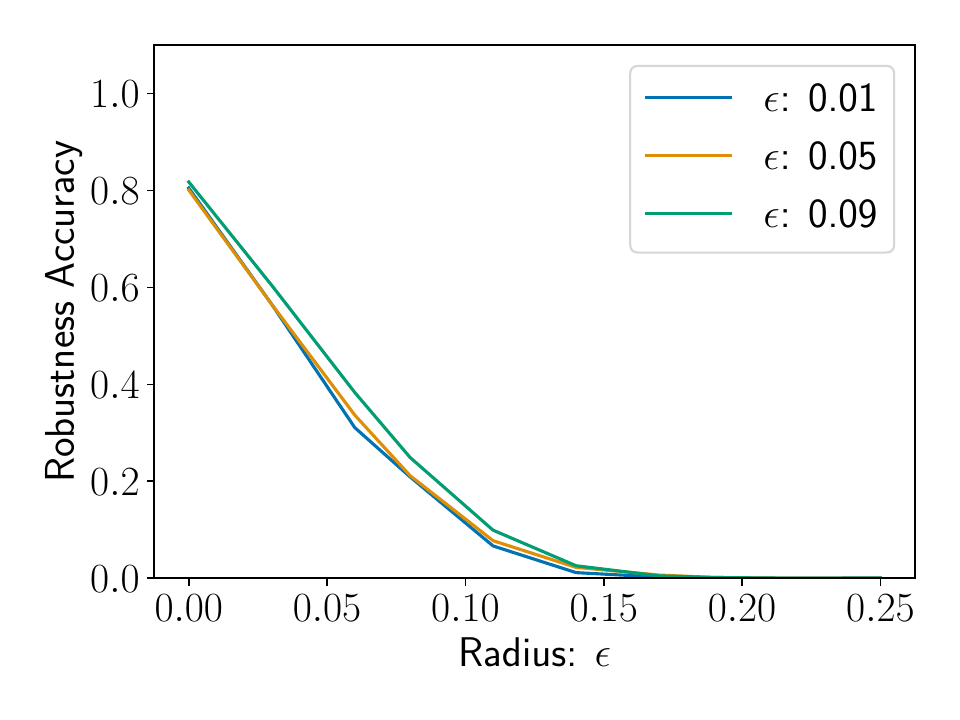}
        \caption{MNIST10: PGD}
    \end{subfigure}
    \\
    \begin{subfigure}{0.3\linewidth}
        \includegraphics[width=\textwidth]{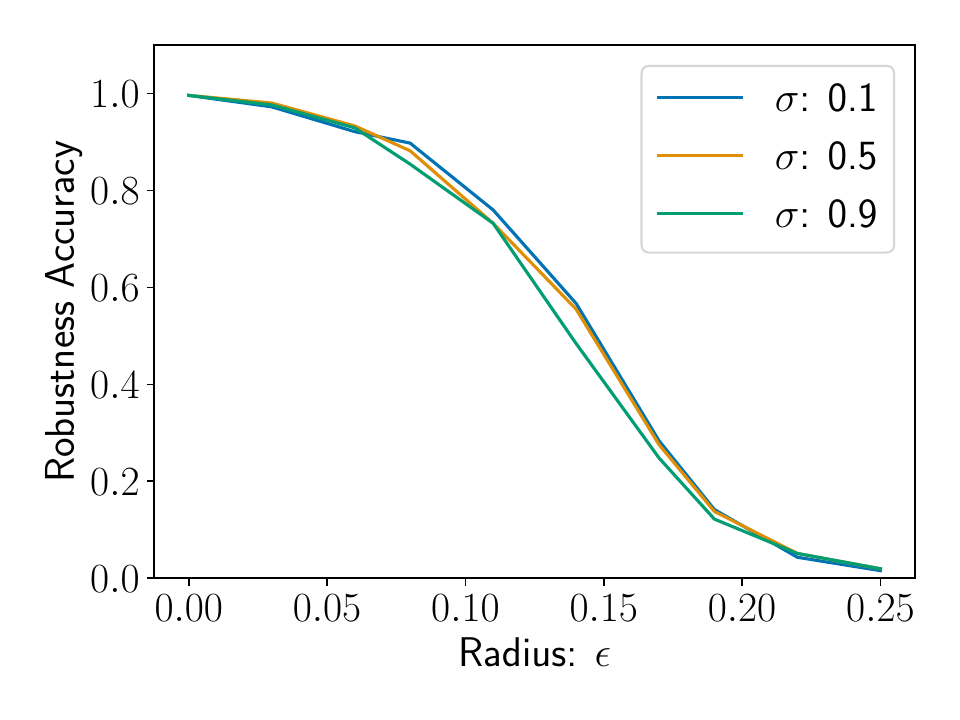}
        \caption{MNIST2: RS }
    \end{subfigure}
    \begin{subfigure}{0.3\linewidth}
        \includegraphics[width=\textwidth]{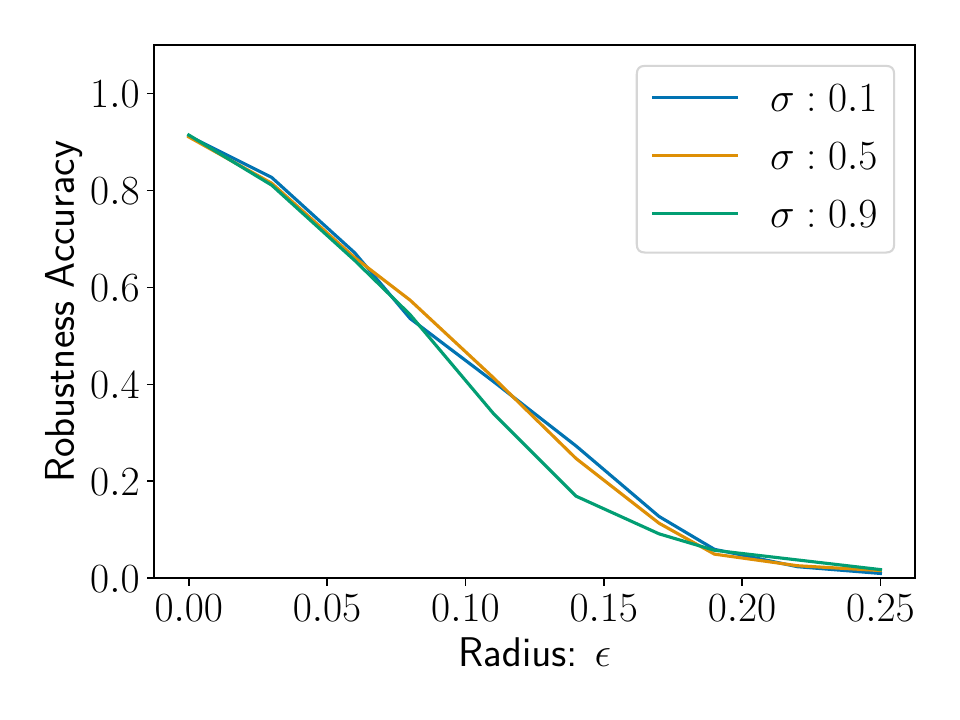}
        \caption{MNIST4: RS}
    \end{subfigure}
    \begin{subfigure}{0.3\linewidth}
        \includegraphics[width=\textwidth]{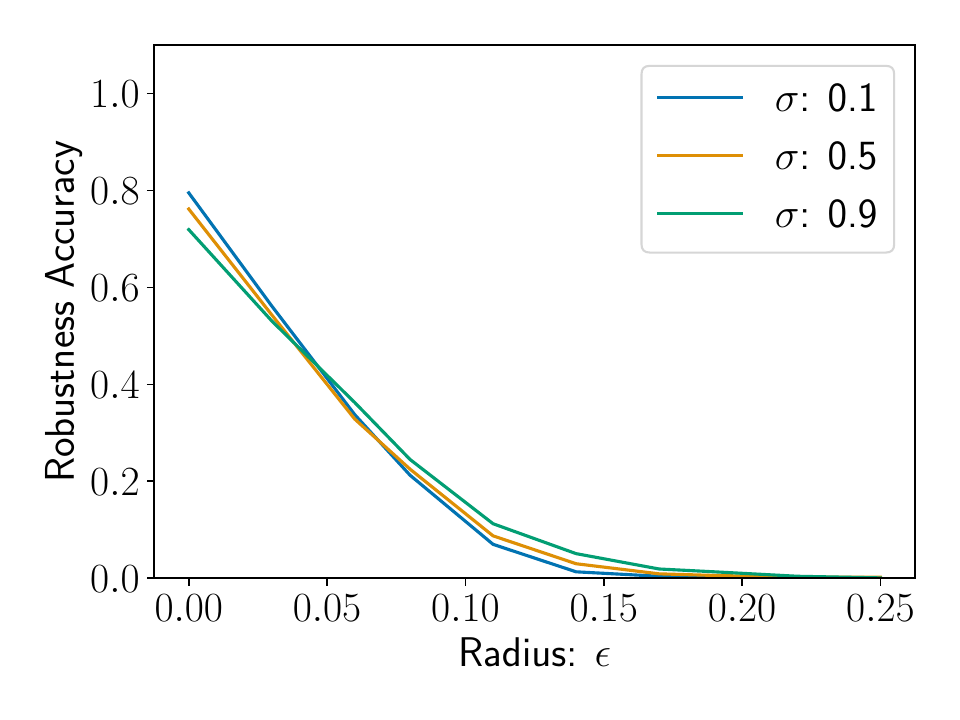}
        \caption{MNIST10: RS}
    \end{subfigure}
    \caption{Comparison of the two methods randomized smoothing (RS) and adversarial training (PGD) for a set of parameters on the QNN. Adversarial training was conducted with PGD($L_\infty$, $T=50$) for the Iris classifier, PGD($L_\infty$, $T=3$) for all MNIST classifiers}
    \label{fig:adversarial_training}
\end{figure*}

Here, we observe the effect of adversarial training and randomized smoothing on the Iris dataset (see \autoref{fig:iris_adversarial}). Overall, we find that adversarial training itself has no influence for the selected epsilon values. 
In contrast, for a large radius in randomized smoothing, the adversarial accuracy drops faster initially but remains more resilient for larger epsilon values.
As a next step, we conduct the same experiments on the MNIST2, MNIST4, and MNIST10 datasets, to examine the influence of adversarial training and randomized smoothing for a larger set of classes. 
Our results are visualized in \autoref{fig:adversarial_training}. 
We observe that neither adversarial training nor randomized smoothing significantly enhances the robustness of the QNN. Additionally, a similar behavior to that seen with depolarizing noise is evident, specifically that the networks tend to become less robust with larger classes. This trend parallels the fact that the testing accuracy drops by approximately 10\% from MNIST2 to MNIST4 and around 10\% from MNIST4 to MNIST10. Interestingly, randomized smoothing and adversarial training do not tend to worsen the robustness accuracy. 

\subsection{Effect of Depolarization Noise and Adversarial Training}
\begin{figure*}[]
    \centering
    \begin{subfigure}{\textwidth}
        \centering
        \includegraphics[width=0.3\textwidth]{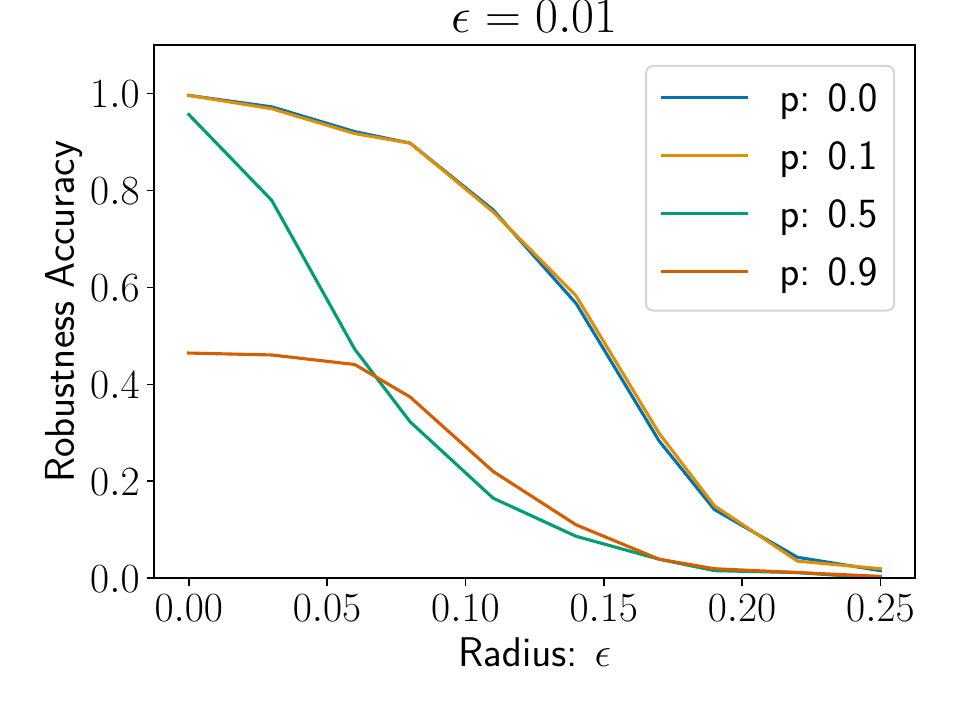}
        \includegraphics[width=0.3\textwidth]{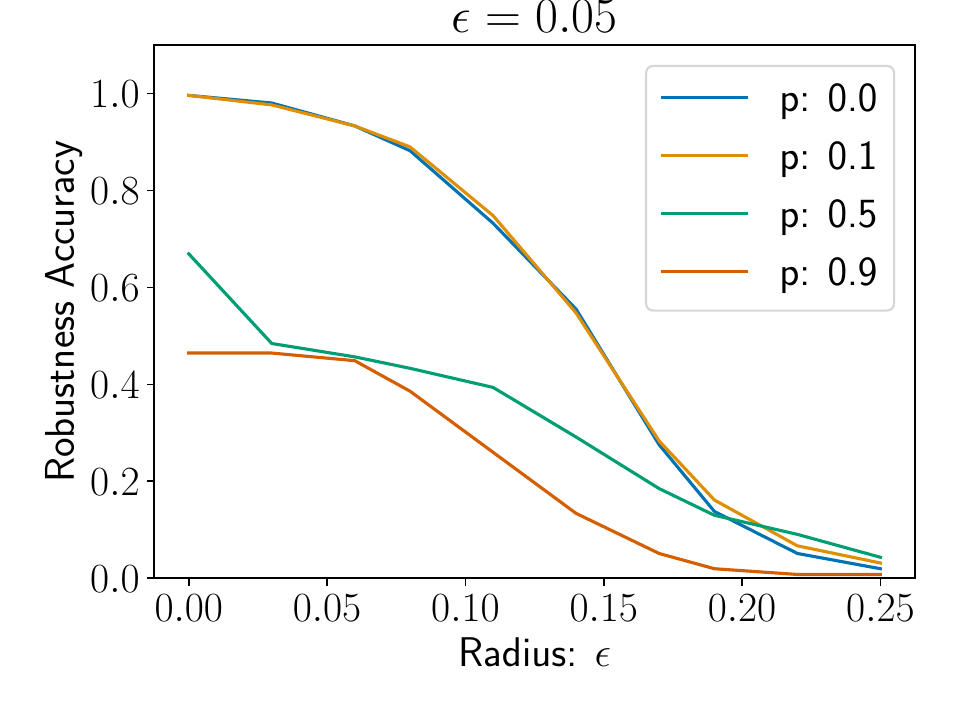}
        \includegraphics[width=0.3\textwidth]{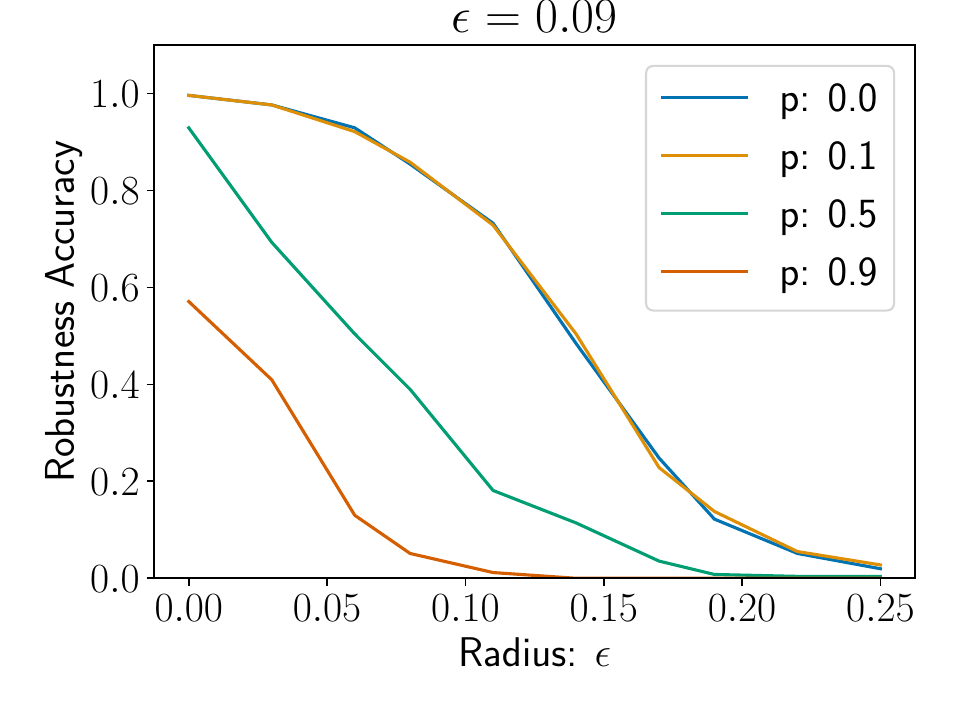}
        \caption{MNIST2: Adversarial Training}
    \end{subfigure}
    \begin{subfigure}{\textwidth}
        \centering
        \includegraphics[width=0.3\textwidth]{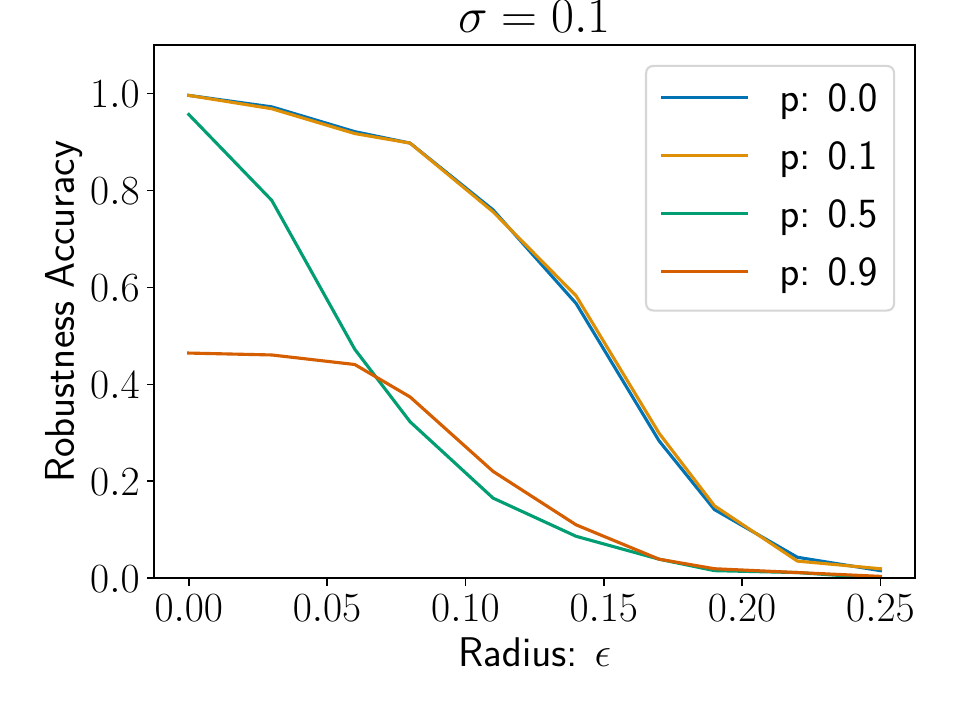}
        \includegraphics[width=0.3\textwidth]{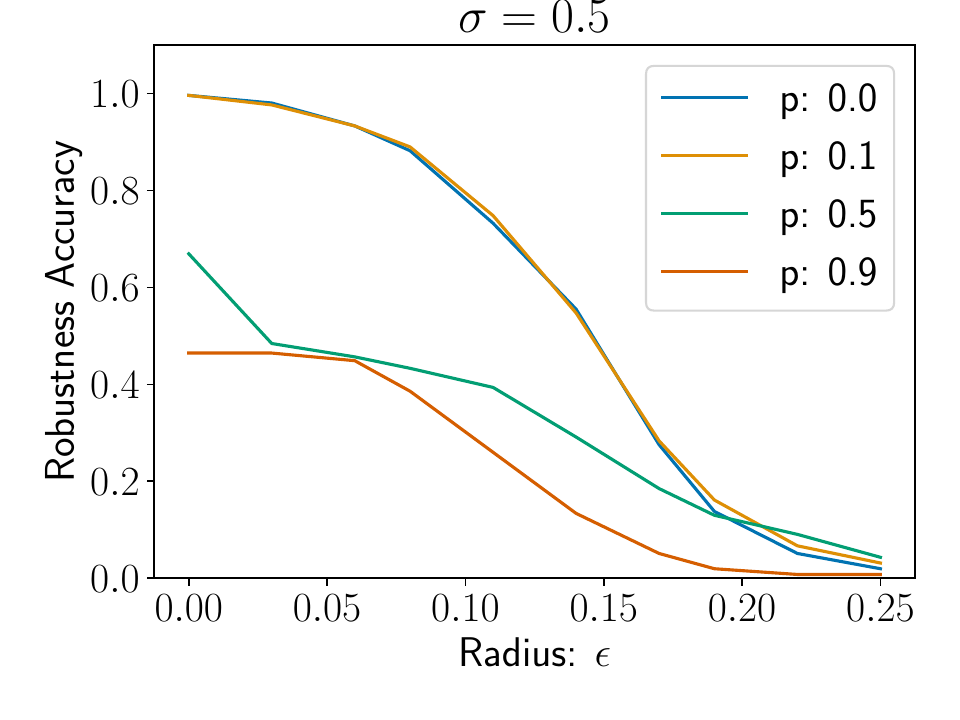}
        \includegraphics[width=0.3\textwidth]{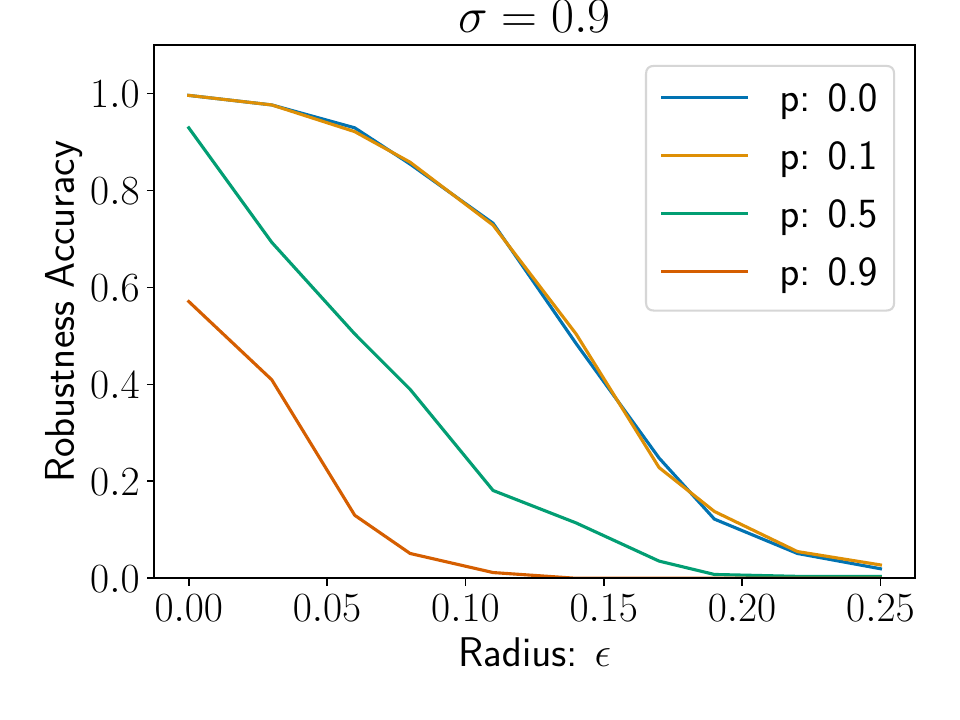}
        \caption{MNIST2: Randomized Smoothing}
    \end{subfigure}
    \caption{Comparison of the adversarially trained QNNs with respect to different depolarization noise.}
    \label{fig:adversarial_training_dep_noise}
\end{figure*}

Since adversarial training and the effect of depolarization noise behave similarly, we explore whether adding depolarization noise \textit{on top} of an adversarial training procedure can improve the general robustness of the network. Therefore, we closely examine the MNIST2 dataset and enhance the adversarially trained QNN circuits with depolarization noise while attacking them. Our results are observed in \autoref{fig:adversarial_training_dep_noise}. Overall, we find that adding depolarization noise on top of adversarial training does not provide improvements for the MNIST2 dataset. 
A similar picture emerges for other datasets when examining the certified trace distances on the test set in \autoref{tab:model_accuracy}.

\section{Conclusion}
Evaluating the robustness of QNNs, we addressed three primary research questions. We assessed the impact of depolarization noise on a quantum classifier. 
For the binary classifiers, we were able to observe that adding depolarizing noise might improve the robustness accuracy, nevertheless only for larger epsilon values and typically after the classifier was perturbed below the $50\%$ mark. We doubt that this behavior is helpful in practice, since one would like to provide a classifier that is more robust for smaller epsilon values.
Next, it was observed that introducing more classes to the QNN decreases both training and test accuracy. Furthermore, the overall robustness of the QNN diminishes and is not improved by depolarization noise. This trend is particularly evident when examining the certifiable distances \(\tau_D\); for all MNIST datasets, \(\tau_D\) decreases with an increase in depolarizing noise and even tends towards negative values when \( p = 0.9 \) for MNIST4 and MNIST10.

Exploring alternative methods for adversarial training of the network also failed to yield benefits, as did the addition of depolarization noise to adversarially trained networks.
\smallskip

It is important to note that \citet{Du_2021} asserted that the certifiable distance depends solely on the overall depolarizing noise, the dimensions, and the ratio of the true class probability to the probability of the subsequent class. Our work demonstrates that increasing depolarizing noise in a quantum channel negatively impacts both this certifiable distance and the adversarial accuracy in practice. Consequently, future research should focus on obtaining weights that enhance class separation in the QNN context. This is crucial because neither the number of measurements nor the depolarizing noise in a quantum device can be controlled. This premise initially motivated our exploration of certifiable distance. In this regard, it would be interesting to build on the work of \citet{liao_robust_practice}, to find more suitable architectures than the one used in this work.


\section*{Acknowledgement}
The research is supported by the Bavarian Ministry of Economic Affairs, Regional Development and Energy with funds from the Hightech Agenda Bayern.

\bibliographystyle{plainnat}
\bibliography{references}

\end{document}